\documentclass[reprint,aps,prb,superscriptaddress,floatfix]{revtex4-2}

\usepackage{amsmath,amsfonts, amssymb, dsfont}
\usepackage{yfonts}
\usepackage{bm}
\usepackage{graphicx}
\usepackage{verbatim}
\usepackage{hyperref}
\usepackage{xcolor}
\usepackage{enumerate}
\usepackage{tikz}
\usepackage{bbold}
\usepackage{booktabs}
\usepackage{multirow}
\usepackage{booktabs}
\usepackage[utf8]{inputenc}
\usepackage[T1]{fontenc}
\usepackage{xcolor}
\usepackage{chemformula}
\usepackage{natbib}
\usepackage{tikz}

\usepackage{diagbox}
\usepackage{tabularx}
\usepackage{float}

\IfFileExists{newtxtext.sty}
{\usepackage{newtxtext,newtxmath}}
{\IfFileExists{stix.sty}
{\usepackage{stix}}
{\IfFileExists{mathptmx.sty}
{\usepackage{mathptmx}}{} } }

\usepackage{textcomp}
\usepackage{bm}

\usepackage{booktabs,siunitx}

\pdfoutput=1
\usepackage{color}
\definecolor{LinkColor}{rgb}{0.256,0.439,0.588}
\usepackage{hyperref}

\hypersetup{
colorlinks=true,
citecolor=LinkColor,
linkcolor=LinkColor,
urlcolor=LinkColor
}

\begin{document}

\title{Universal Scaling Functions of the Gr{\"u}neisen Ratio near Quantum Critical Points}

\author{Xuan Zhou}
\thanks{These authors contributed equally to this work.}
\affiliation{State Key Laboratory of Surface Physics, Fudan University, Shanghai 200438, China}
\affiliation{Center for Field Theory and Particle Physics, Department of Physics, Fudan University, Shanghai 200433, China}

\author{Enze Lv}
\thanks{These authors contributed equally to this work.}
\affiliation{Institute of Theoretical Physics, Chinese Academy of Sciences, 
	Beijing 100190, China}
\affiliation{School of Physical Sciences, University of Chinese Academy of 
Sciences, Beijing 100049, China}

\author{Wei Li}
\email{w.li@itp.ac.cn}
\affiliation{Institute of Theoretical Physics, Chinese Academy of Sciences, 
	Beijing 100190, China}
\affiliation{School of Physical Sciences, University of Chinese Academy of 
Sciences, Beijing 100049, China}

\author{Yang Qi}
\email{qiyang@fudan.edu.cn}
\affiliation{State Key Laboratory of Surface Physics, Fudan University, Shanghai 200438, China}
\affiliation{Center for Field Theory and Particle Physics, Department of Physics, Fudan University, Shanghai 200433, China}
\affiliation{Collaborative Innovation Center of Advanced Microstructures, Nanjing 210093, China}

\date{\today}

\begin{abstract}
The Gr\"uneisen ratio, defined as $\Gamma_g \equiv (1/T) (\partial T/\partial g)_S$, serves as a highly sensitive probe for detecting quantum critical points (QCPs) driven by an external feild $g$ and for characterizing the magnetocaloric effect (MCE). Near a QCP, the Gr\"uneisen ratio displays a universal divergence which is governed by a universality-class-dependent scaling function stemming from the scale invariance. In this work, we systematically investigate the universal scaling functions of Gr\"uneisen ratio in both one-dimensional (1D) and two-dimensional (2D) quantum spin systems, including the transverse-field Ising model, the spin-1/2 Heisenberg model, the quantum $q$-state Potts model ($q=3,4$) and the $J_1$-$J_2$ columnar dimer model. Our approach employs the thermal tensor-network method for infinite-size 1D systems and the stochastic series expansion quantum Monte Carlo (SSE QMC) simulations for 2D systems, enabling precise calculations of the Gr\"uneisen ratio near QCPs. Through data collapse analysis, we extract the corresponding scaling functions, which establish quantitative frameworks to interpret magnetocaloric experiments and guide the development of ultralow-temperature refrigeration. 
\end{abstract}

\maketitle

\section{Introduction}
\begin{figure}[htp!]
    \centering
    \includegraphics[width=1\columnwidth]{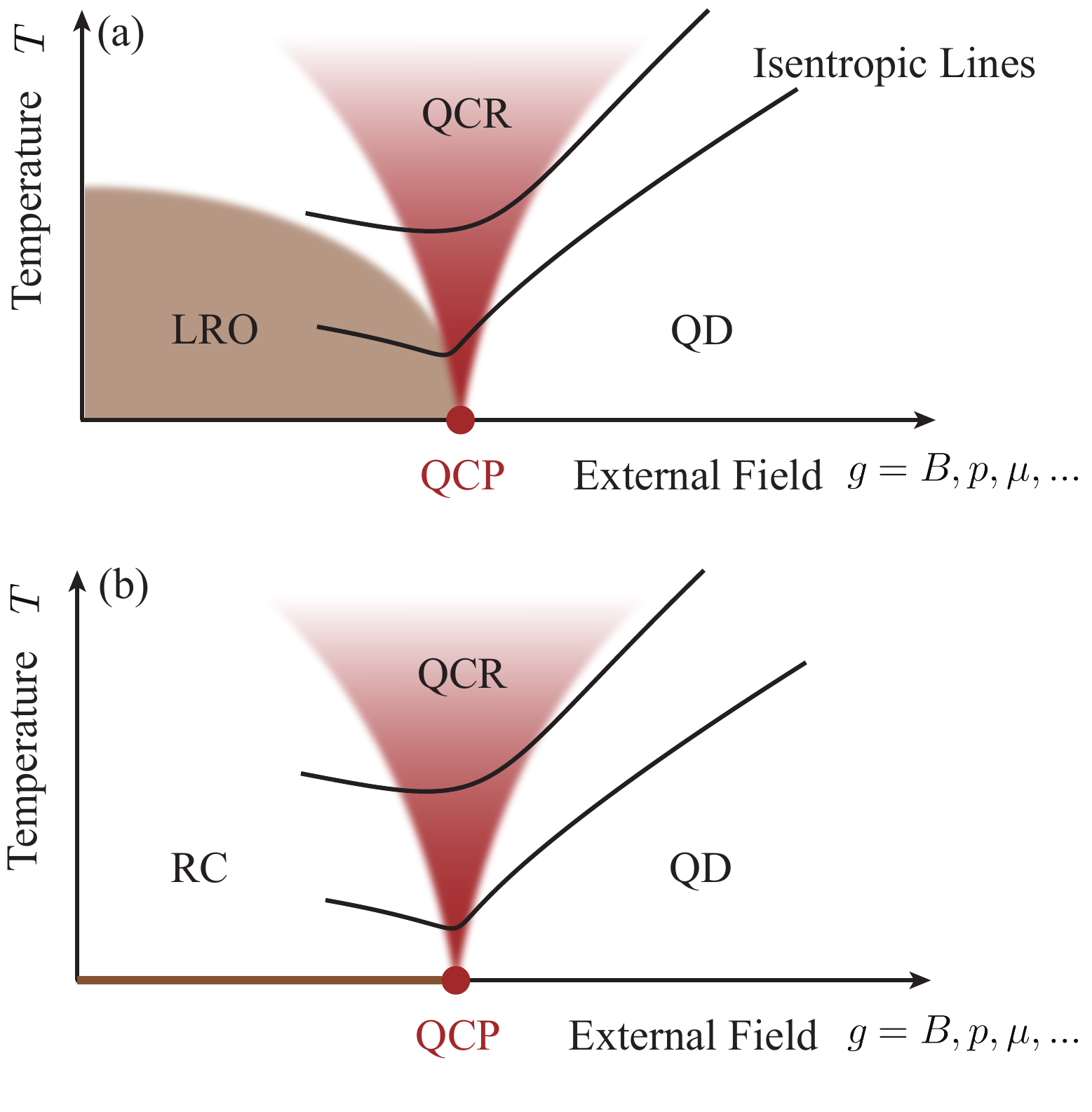}
    \caption{\textbf{Two possible schematic phase diagrams.} The red dot represents a quantum critical point (QCP), while the red cone is the finite-temperature quantum critical regime (QCR). The black lines represent the schematic isentropic lines in the $g$-$T$ plane, where the external field $g$ could be magnetic field $B$, pressure $p$, chemical potential $\mu$ and so on. (a) The system exhibits long-range order (LRO) at finite temperatures and the QCR separates the LRO and the quantum disordered (QD) regime. (b) The LRO vanishes at finite temperatures. The QCR separates the renormalized classical (RC) regime and the QD regime. }
    \label{fg:phase_diagram}
\end{figure}

Quantum phase transition (QPT), which occurs at zero temperature upon variation of a nonthermal control parameter, is a topic of great research interest in condensed matter physics~\cite{sachdev_2011,Matthias_Vojta_2003}. When the QPT is continuous, it is marked by a quantum critical point (QCP), where the energy scale vanishes and the system becomes scale-invariant. At finite temperatures, as shown in Fig.~\ref{fg:phase_diagram}, the interplay between thermal and quantum fluctuations gives rise to a quantum critical regime (QCR) that hosts universal scaling laws and other emergent quantum critical phenomena~\cite{Sachdev2000,Coleman2005}, e.g., non-Fermi liquid~\cite{RevModPhys.73.797,RevModPhys.79.1015} and unconventional superconductivity~\cite{vanderMarel2003QuantumCB}. 

\begin{table*}[htbp]
    \centering
    \renewcommand{\arraystretch}{1.4}
    \begin{tabular}{ccccccc}
    \toprule
     Universality class & Material & Effective model & Magnetic structure & Critical field [T] & Field direction & Reference \\
     \hline
     \multirow{3}{*}{(1+1)D Ising}& \ch{CoNb_2O_6} & FM Ising & 1D chain & $B_{c}\simeq 5.25$ & $B \parallel b$ & \cite{coldea2010,kinross2014,liang2015,morris2021,xu2022,woodland2023} \\
     & \ch{BaCo_2V_2O_8} & AFM Ising & 1D chain & $B_{c}\simeq 4.7$ & $B \parallel b$ & \cite{Wang2018,faure2018,zou2021,Wang2024Nature} \\
     & \ch{SrCo_2V_2O_8} & AFM Ising & 1D chain & $B_{c}\simeq 7.7$ & $B \parallel a$ & \cite{He2006,Wang2018Nature,cui2019,Zou_2019} \\
    \hline
    (1+1)D Potts & Rydberg atoms & long-range Ising & 1D chain & --- & --- & \cite{Bernien2017,Maceira2022} \\ 
    \hline
    \multirow{4}{*}{(1+2)D BEC} & \ch{CuPzN} & AFM Heisenberg & 1D chain & $B_{c}\simeq 13$ & $B \parallel b$ &\cite{doi:10.1126/sciadv.aao3773} \\ 
      & CuP & AFM Heisenberg & 1D chain & $B_{c}\simeq 4.09$  &  $B \parallel b$ & \cite{doi:10.1073/pnas.1017047108}\\ 
      & CuSO$_4\cdot$5H$_2$O & AFM Heisenberg & 1D chain & $B_c^{1D}\simeq 3.9$ & --- & \cite{xiang2025} \\
      & \ch{(C_5H_{12}N)_2CuCl_4} & AFM Heisenberg & two-leg ladder & $B_{c1}\simeq 1.73, B_{c2}\simeq 4.4$ & $B \parallel a$ & \cite{CCCMCE2014}\\ 
      \hline
     (2+1)D Ising
      & \ch{K_2Co(SeO_3)_2} & AFM easy-axis Heisenberg & 2D triangular lattice & --- & $B \perp c$ & \cite{RCSO2020,zhu2024continuum,chen2024phase}  \\
      \hline
     (2+1)D O(3) & ultracold atoms & Fermi-Hubbard & 2D square lattice & --- & --- & \cite{Xu2025}\\
       \hline
       \multirow{4}{*}{(2+2)D BEC}  & \ch{Na_2BaCo(PO_4)_2}  & AFM easy-axis Heisenberg & 2D triangular lattice & $B_{c1}\simeq 0.36, B_{c2}\simeq 1.14$  &  $B \parallel c$  & \cite{Zhong2019,LiN2020,Wu2022PNAS,Gao_2022,xiang2024giant,Gao2024} \\
       & \ch{YbCl_3} & AFM Heisenberg & 2D honeycomb lattice & $B_c\simeq 5.9$ & $B\parallel a$ & \cite{Yosuke2024} \\
       & \ch{Cu(DCOO)2*}4\ch{D2O} & AFM Heisenberg & 2D square lattice & --- & --- & \cite{Dalla_Piazza_2014,Christensen_2007,PhysRevLett.87.037202} \\
       & \ch{K_2Co(SeO_3)_2} & AFM easy-axis Heisenberg & 2D triangular lattice & $B_{c1}\simeq 0.8$, $B_{c2}\simeq 19$ & $B \parallel c$ & \cite{RCSO2020,zhu2024continuum,chen2024phase} \\
        \hline
       \multirow{3}{*}{(3+2)D BEC} & CuSO$_4\cdot$5H$_2$O & AFM Heisenberg & 3D lattice & $B_c^{3D}\simeq 4.15$ & --- & \cite{xiang2025} \\
       & NiCl$_2$-4SC(NH$_2$)$_2$ & AFM Heisenberg ($S=1$) & 3D lattice & $B_{c1}\simeq 2.1$, $B_{c2}\simeq 2.6$ & $B\parallel c$ & \cite{Zapf2006} \\
       & \ch{TlCuCl_3} & --- & --- & --- & --- & \cite{Ruegg2003,PhysRevLett.95.017205} \\
      
    \bottomrule
    \end{tabular}
    \caption{Material realizations of typical quantum universality classes.}
    \label{tab:Materials}
\end{table*}

Conventionally, QCP and its associated QCR can be detected experimentally by measuring the order parameter~\cite{Xibo2012}. However, this approach faces limitations since some QCPs lack a well-defined and clearly identifiable order parameter~\cite{xiang2024giant,Li2024TopoCooling}. Therefore, the development of more powerful methods for probing quantum criticality remains essential. In fact, the Gr\"uneisen ratio $\Gamma_g\equiv (1/T)\left(\partial T/{\partial g}\right)_S$~\cite{https://doi.org/10.1002/andp.19123441202}, which characterizes the adiabatic response of a system to an external field $g$, is emerging as a sensitive probe for investigating quantum criticality in both theoretical and experimental studies~\cite{PhysRevB.72.205129,doi:10.1126/sciadv.aao3773,Gegenwart_2016}. 
Within the QCR, the isentropic lines reach their minimum temperatures (see Fig.~\ref{fg:phase_diagram}), corresponding to the sign changes of the Gr\"uneisen ratio~\cite{PhysRevB.72.205129}. Due to the presence of collective low-energy excitations that carry significant thermal entropy, quantum critical fluctuations give rise to a cooling effect during adiabatic demagnetization processes, which has been utilized for solid-state refrigeration in the sub-Kelvin regime~\cite{xiang2024giant,xiang2025,Gschneidner_2005,Lang_2012}. On the other hand, renormalization group theory and scaling analysis indicate that the Gr{\"u}neisen ratio displays universal divergence upon approaching the QCP~\cite{PhysRevLett.91.066404,Lv2024}, and its efficacy as a probe for QCPs has been observed in various realistic quantum systems, including spin-chain compounds with field-induced QCPs~\cite{coldea2010,kinross2014,liang2015,morris2021,xu2022,woodland2023,Wang2018,faure2018,zou2021,Wang2024Nature,He2006,Wang2018Nature,cui2019,Zou_2019,doi:10.1126/sciadv.aao3773} and heavy-fermion metals~\cite{article,PhysRevLett.93.096402,PhysRevLett.102.066401,Gegenwart_2016,PhysRevLett.91.066405}. Consequently, the Gr\"uneisen ratio with its scaling functions is an effective method to detect the global constraints such as symmetry, dimensionality, and self-duality of the system~\cite{xiang2025}.

In recent years, with the development of experimental technologies, researchers have been able to measure the Gr\"uneisen ratio precisely and extract its scaling function~\cite{doi:10.1126/sciadv.aao3773,xiang2025}. As summarized in Table~\ref{tab:Materials}, a number of representative materials that host QCPs and have been experimentally investigated, showing potential significance for magnetocaloric measurements. Notably, although the magnetocaloric effect (MCE) near QCPs can be experimentally measured, theoretical calculations remain relatively scarce. It is therefore significant to theoretically calculate the Gr\"uneisen ratio and its corresponding scaling functions across various QCPs to provide a priori guidance and predictions for future experiments. 

In this paper, we present the calculations of the Gr{\"u}neisen ratio for spin models using thermal tensor network method and stochastic series expansion quantum Monte Carlo (SSE QMC) method. We systematically investigate various QCPs across multiple universality classes, including (1+1)D Ising, (2+1)D Ising, (1+1)D 3-state Potts, (1+1)D 4-state Potts, (2+1)D XY, (2+1)D O(3) and (1+2)D Bose-Einstein condensation (BEC). By performing large-size simulations at low temperatures, we determine the field dependence of the Gr\"uneisen ratio in the vicinity of these QCPs. Based on scaling hypothesis, we perform data collapse to extract the universal scaling functions of Gr{\"u}neisen ratio. 

The remainder of this paper is organized as follows. In Section~\ref{section2}, we give the universal form of the Gr{\"u}neisen ratio. In Section~\ref{section3}, we introduce the spin models and associated material realizations. In Section~\ref{section4}, we show the results for the universal scaling function and explain the anomalous behavior of the Gr{\"u}neisen ratio near the (2+1)D O(3) QCP. Section \ref{section5} includes the discussions and conclusions of the paper.

\section{Scaling function of the Gr{\"u}neisen ratio}
\label{section2}
Driven by pressure, the Gr{\"u}neisen ratio $\Gamma$ is defined as the ratio between thermal expansion $\alpha$ and specific heat $c_p$, i.e., $\Gamma = \alpha/c_p$. Generically, induced by a nonthermal parameter $g$, the Gr{\"u}neisen ratio
 can be written as,
 \begin{equation}
 \Gamma_g=\frac{\alpha_g}{c_g}=-\frac{1}{T}\frac{(\partial S/\partial g)_T}{(\partial S/\partial T)_g} = \frac{1}{T}\left(\frac{\partial T}{\partial g}\right)_S,
 \label{gamma}
 \end{equation}
 where $S$ is the thermal entropy and $c_g$ is the specific heat at external field $g$. Given the Maxwell relations, the generalized ``thermal expansion'' $\alpha_g$ reflects the entropy change induced by the quantum fluctuation $g$, and the Gr\"uneisen ratio $\Gamma_g$ characterizes the temperature change under adiabatic processes. For example, when $g$ is the magnetic field, $\Gamma_g$ describes the MCE, which is mainly discussed in our work.

Near a QCP, thermodynamic properties exhibit universal forms due to the divergent correlation length. Particularly, the Gr\"uneisen ratio takes the following forms in the vicinity of QCP (see Appendix~\ref{append:scaling} for details): 
\begin{equation}
\Gamma(T,g)= \tilde{T}^{-1/z\nu}/g_0\cdot \phi\left(x\right),
\label{fT}
\end{equation}
\begin{equation}
\Gamma(T,g)= \tilde{g}^{-1}/g_0\cdot \tilde{\phi}\left(x\right),
\label{fg}
\end{equation}
where $\nu$ is the critical exponent of correlation length, $z$ is the dynamic exponent, and $d$ is the spatial dimension. Note that $\tilde{g}\equiv(g-g_c)/g_0$ with $g_0$ the characteristic external field measures the distance to the QCP at $g_c$ and $\tilde{T} \equiv T/T_0$ is the rescaled temperature with $T_0$ the characteristic temperature of the system, while $T_0$ and $g_0$ are nonuniversal constants. In this paper, we set $T_0=g_0=g_c$ for simplicity. Importantly, the universal scaling functions $\phi(x)$ and $\tilde{\phi}(x)\equiv x\phi(x)$, with $x\equiv \tilde{g}\tilde{T}^{-1/z\nu}$, are the primary focus of this work. If we calculate the $\phi(x)$ or $\tilde{\phi}(x)$, the Gr\"uneisen ratio near the QCP can be determined completely. After aligning the energy scales $T_0$ and $g_0$, magnetocaloric results will follow the corresponding universal scaling function, independent of the specific Hamiltonian and other details of realistic materials. 

\section{Model Hamiltonian}
\label{section3}
We consider typical quantum spin models on 1D chain and 2D square lattice which host QCPs induced by external field $g$. The first example is transverse-field Ising (TFI) model, described by
\begin{equation}
H_{\rm TFI}=-J\sum_{\langle i j \rangle}\sigma_i^z \sigma_j^z -\Delta\sum_i\sigma_i^x,
\label{TFI}
\end{equation}
where $\langle i j \rangle$ represents the nearest-neighbor bond, $\sigma^z$ and  $\sigma^x$ are Pauli matrices. We set the Ising coupling $J=1$ as the energy scale. The transverse field $\Delta$ drives a second-order QPT at the critical field $\Delta_c$. For $\Delta<\Delta_c$, the system breaks the $\mathbb{Z}_2$ symmetry and develops ferromagnetic order in the ground state, while for $\Delta>\Delta_c$, it becomes quantum disordered. In 1D TFI model, a self-dual QCP appears at $\Delta_c=1$, which belongs to the (1+1)D Ising universality class. There exists a lot of Ising magnets that exhibit the (1+1)D Ising quantum critical behaviors in experiments, including \ch{CoNb_2O_6}~\cite{coldea2010,kinross2014,liang2015,morris2021,xu2022,woodland2023}, \ch{BaCo_2V_2O_8}~\cite{Wang2018,faure2018,zou2021,Wang2024Nature} and \ch{SrCo_2V_2O_8}~\cite{He2006,Wang2018Nature,cui2019,Zou_2019}. On the square lattice, there exists a QCP at $\Delta_c=3.04438(2)$ \cite{PhysRevE.66.066110} which is in the (2+1)D Ising universality class. Triangular lattice antiferromagnet \ch{K_2Co(SeO_3)_2} has recently emerged as a strong candidate for realizing (2+1)D Ising QCP~\cite{RCSO2020,zhu2024continuum,chen2024phase}. 

Secondly, we study the 1D quantum $q$-state Potts model, whose Hamiltonian, possessing $S_q$ symmetry, is given by 
\begin{equation}
    H_{\rm Potts} = -J\sum_{\langle i j \rangle} \sum_{k=1}^{q-1} \Omega_i^k \Omega_j^{q-k} - \theta \sum_i \sum_{k=1}^{q-1}\Lambda_i^k,
    \label{Potts}
\end{equation}
where $J=1$ is the FM coupling and $\theta$ represents transverse field. $\Omega$ is a diagonal matrix ${diag}\left(1,\omega,\omega^2,...,\omega^{q-1}\right)$ with $\omega = e^{\frac{2\pi i}{q}}$ and $\Lambda$ is an off-diagonal matrix $\left(0,{\mathbb I}_{(q-1)\times (q-1)};1,0\right)$. Particularly, for the $q=2$ case, Eq.~(\ref{Potts}) reduces to the transverse-field Ising model shown in Eq.~(\ref{TFI}). At the critical field $\theta_c=1$, there exists a self-dual QCP belonging to the (1+1)D $q$-state Potts universality class. These $q$-state QCPs can be realized in Rydberg atom chains by tuning the Rabi frequency and the laser detuning~\cite{Bernien2017,Maceira2022}. 

\begin{figure}[htbp]
    \centering
    \includegraphics[width=1\columnwidth]{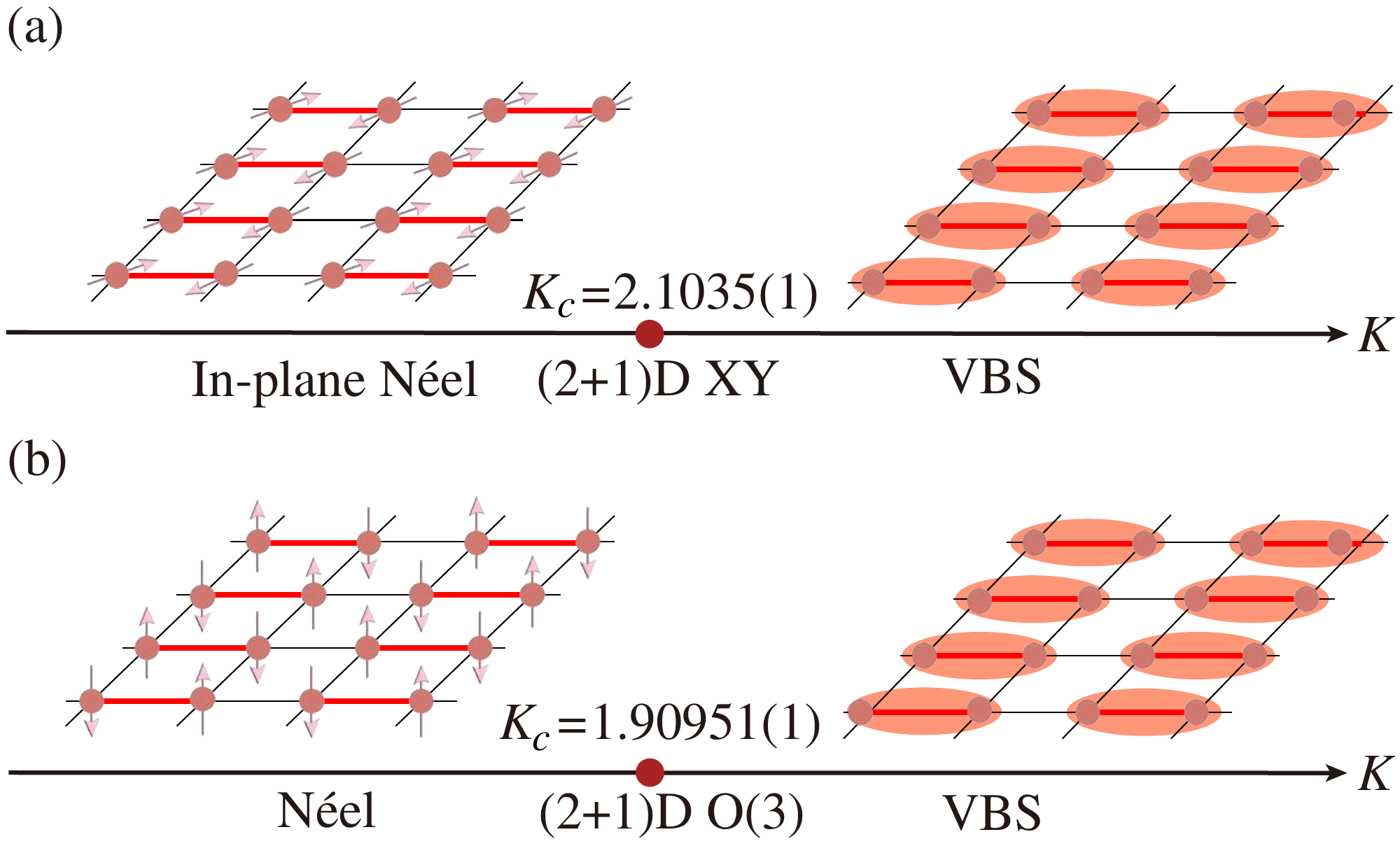}
    \caption{\textbf{Phase diagram of $J_1$-$J_2$ columnar dimer model.} (a) Easy-plane anisotropic case $\lambda=0.9$. (b) Isotropic case $\lambda=1$.}
    \label{fg:J1J2_phase_diagram}
\end{figure}

The third example is the 1D spin-1/2 Heisenberg antiferromagnetic (HAF) model in a uniform magnetic field $h$. The Hamiltonian in standard notation reads 
\begin{equation}
H_{\rm HAF}=J\sum_{\langle i j\rangle} \left(S_i^xS_j^x+S_i^yS_j^y+ S_i^zS_j^z\right)-h\sum_i S_i^z,
\end{equation} 
where $J=1$ manifests the antiferromagnetic (AFM) coupling. There exists a QCP located at $h_c=2$ where the Bose-Einstein condensation (BEC) of magnons occurs. The QCP, which falls into the (1+2)D BEC universality class, separates the gapless Tomonaga-Luttinger liquid ($h<h_c$) and the gapped paramagnetic phase ($h>h_c$). The (1+2)D BEC transition has been extensively observed and studied in a variety of 1D~\cite{doi:10.1126/sciadv.aao3773,doi:10.1073/pnas.1017047108,xiang2025} and quasi-1D~\cite{CCCMCE2014} quantum magnets, where strong quantum fluctuations and reduced dimensionality stabilize exotic critical behavior. For the 2D square-lattice case, a finite-temperature Berezinskii-Kosterlitz-Thouless (BKT) phase transition occurs for $h<h_c$~\cite{PhysRevLett.130.086704,PhysRevB.68.060402}. The magnetic field $h$ induces a QCP at $h_c$, driving the system from a quantum disordered phase ($h>h_c$) to an in-plane AFM phase ($h<h_c$) upon breaking the U(1) symmetry. The critical field $h_c=4$ is determined by quantum Monte Carlo method based on the worm algorithm~\cite{PhysRevB.68.060402}. This QCP falls into the (2+2)D BEC universality class, which reaches the upper critical dimension and its critical behaviors can be understood by the mean-field theory. Besides square lattice antiferromagnet \ch{Cu(DCOO)2*}4\ch{D2O}~\cite{Dalla_Piazza_2014,Christensen_2007,PhysRevLett.87.037202}, such (2+2)D BEC quantum criticality can be realized in other certain quantum materials, such as \ch{Na_2BaCo(PO_4)_2}~\cite{Zhong2019,LiN2020,Wu2022PNAS,Gao_2022,xiang2024giant,Gao2024}, \ch{YbCl_3}~\cite{Yosuke2024} and \ch{K_2Co(SeO_3)_2}~\cite{RCSO2020,zhu2024continuum,chen2024phase}.

\begin{figure*}[htbp]
    \centering
     \includegraphics[width=2\columnwidth]{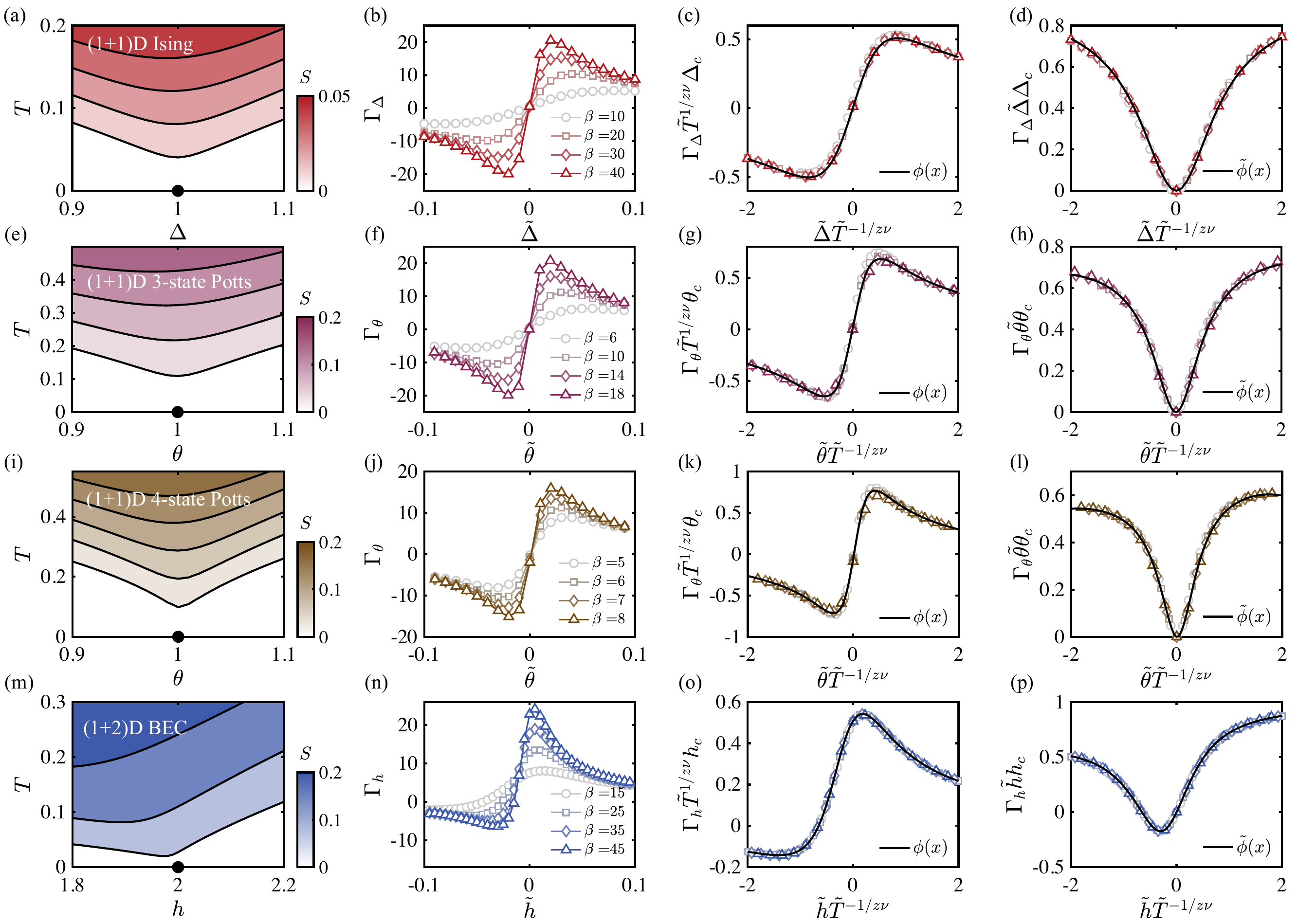}
    \caption{\textbf{Gr{\"u}neisen ratio $\Gamma_g$ calculated for 1D systems in the thermodynamic limit.} (a,e,i,m) Isentropic lines for the 1D TFI model, quantum 3-state Potts model, quantum 4-state Potts model and HAF model. The black solid dots represent QCPs, with their corresponding universality classes indicated above. (b,f,j,n) Gr{\"u}neisen ratio as a function of external field for different inverse temperatures $\beta$. (c,g,k,o) The data collapse of $\Gamma_g$ near the QCP according to Eq.~(\ref{fT}). Here we choose $T_0=g_0=g_c$ and use the (1+1)D Ising critical exponent $\nu=1 ,\,z=1$, (1+1)D 3-state Potts critical exponent $\nu=5/6 ,\,z=1$, (1+1)D 4-state Potts critical exponent $\nu=2/3 ,\,z=1$ and (1+2)D BEC critical exponent $\nu=1/2 ,\,z=2$ in the scaling sequentially. (d,h,l,p) The data collapse of $\Gamma_g$ according to Eq.~(\ref{fg}). Black solid lines illustrate the scaling functions $\phi(x)$ and $\tilde{\phi}(x)$. }
    \label{fg:1D_results}
\end{figure*}

Finally, we study the 2D $J_1$-$J_2$ AFM columnar dimer model on the square lattice defined by the Hamiltonian
\begin{equation}
H_{\rm J_1,J_2}=J_1\sum_{\langle i j \rangle} D_{ij} + J_2\sum_{\langle i j \rangle^{\prime}} D_{ij},
\end{equation}
where $J_1,J_2>0$ are the AFM couplings defined on distinct nearest-neighbor bonds $\langle i j \rangle$ and $\langle i j \rangle^{\prime}$, respectively. As illustrated in Fig.~\ref{fg:J1J2_phase_diagram}, the red thick $J_2$ bonds form an alternating pattern along the $x$-direction, while the remaining black thin bonds are $J_1$ type. $D_{ij}=S_i^xS_j^x+S_i^yS_j^y+\lambda S_i^zS_j^z$ is the Heisenberg term with $\lambda$ controlling the anisotropy. For the isotropic case ($\lambda=1$), the coupling ratio $K\equiv J_2/J_1$ drives a second-order QPT between the N\'eel order ($K<K_c$) and the columnar valence-bond solid (VBS) phase ($K>K_c$), as shown in Fig.~\ref{fg:J1J2_phase_diagram}(b). The QCP at $K_c=1.90951(1)$ belongs to the (2+1)D O(3) universality class~\cite{PhysRevB.79.014410,PhysRevLett.121.117202}. For the easy-axis case $\lambda>1$, the system undergoes a (2+1)D Ising phase transition by tuning $K$. While for $0\leq \lambda < 1$, U(1) symmetry breaking occurs at the QCP which belongs to the (2+1)D XY universality class [see Fig.~\ref{fg:J1J2_phase_diagram}(a)]~\cite{zhu2022exoticsurfacebehaviorsinduced,PhysRevB.98.174421}. Here, we use the $J_1$-$J_2$ dimer model to investigate the (2+1)D O(3) QCP at $K_c=1.90951(1)$ with the isotropic case and the (2+1)D XY QCP at $K_c=2.1035(1)$ with $\lambda=0.9$ the easy-plane anisotropic case~\cite{PhysRevB.79.014410,zhu2022exoticsurfacebehaviorsinduced}. As far as we know, the (2+1)D XY QCP and the (2+1)D O(3) QCP have not been observed in real quantum materials, but recent research has employed the (2+1)D O(3) QCP to cool ultracold fermionic atoms to cryogenic temperatures for Hubbard quantum simulation~\cite{Xu2025}.

\begin{figure*}[htp!]
    \centering
    \includegraphics[width=2.0\columnwidth]{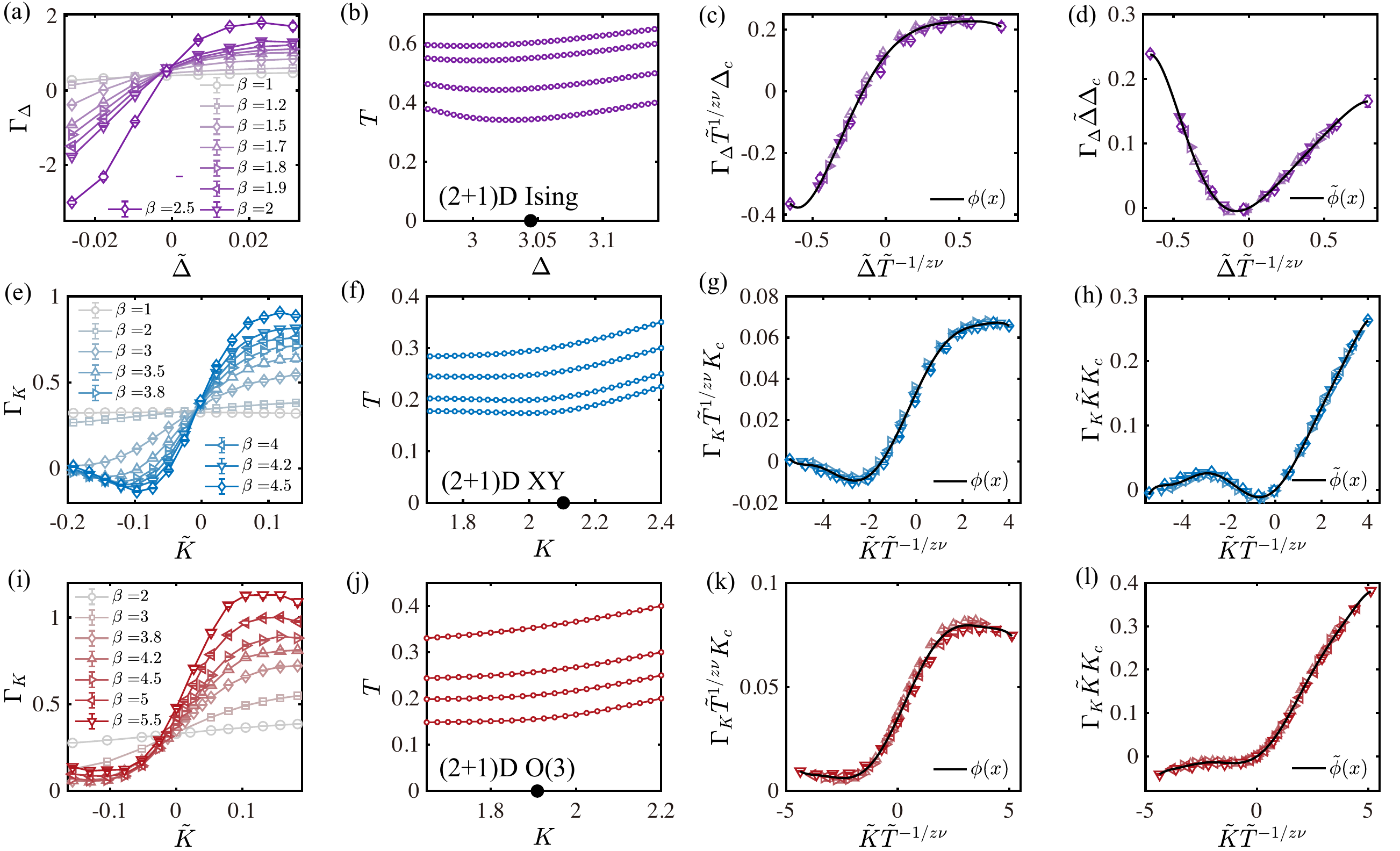}
    \caption{\textbf{Gr{\"u}neisen ratio $\Gamma_g$ calculated for 2D spin models on finite-size lattices with system size $L=24$.} (a,e,i) Gr{\"u}neisen ratio as function of tune parameter for different inverse temperature $\beta$ for the 2D TFI model, 2D easy-plane $J_1$-$J_2$ model model and 2D isotropic $J_1$-$J_2$ model. (b,f,j) Adiabatic demagnetization curve obtained from the integral of the Gr{\"u}neisen ratio for the 2D TFI model, 2D HAF model and 2D $J_1$-$J_2$ model. The black circle represents the corresponding QCP. (c,g,k) The data collapse of $\Gamma_g$ near the QCP according to Eq.~(\ref{fT}). Here we choose $T_0=g_0=g_c$ and use the (2+1)D Ising critical exponent $\nu=0.629971$, (2+1)D XY critical exponent $\nu=0.67175$ and (2+1)D O(3) critical exponent $\nu=0.707$ in the scaling sequentially. (d,h,l) The data collapse of $\Gamma_g$ near the QCP according to Eq.~(\ref{fg}). Black solid lines illustrate the scaling functions $\phi(x)$ and $\tilde{\phi}(x)$. }
    \label{fg:2D_results}
\end{figure*}

\section{Results}
\label{section4}
Using the tensor-network and QMC methods mentioned before, we calculate the Gr\"uneisen ratio and the corresponding scaling functions for these quantum models, as shown in Fig.~\ref{fg:1D_results}-\ref{fg:2D_results}. The polynomial fitting results for the universal functions are also presented in Table~\ref{tab:functions}. For the 1D systems, we directly compute the results in the thermodynamic limit ($L\to \infty$), whereas for the 2D systems, we consider lattice sizes up to $24\times24$ and check the convergence of system size $L$ in the Appendix~\ref{append:convergence}.

First, we present the 1D results, including the TFI model (i.e., the $q=2$ Potts model), $q=3$ quantum Potts model, $q=4$ quantum Potts model and HAF model, as shown in Fig.~\ref{fg:1D_results}. The first three models possess self-dual QCPs, whereas the 1D HAF model does not. Due to the constraint of self-duality, the minimum of the isentropic line occurs nearly above the QCP shown in the Fig.~\ref{fg:1D_results}(a,e,i) and the Gr\"uneisen ratio at the critical field maintains a constant without temperature-dependent divergence [see Fig.~\ref{fg:1D_results}(b,f,j)]. To discuss the quantum critical Gr\"uneisen ratio with self duality in detail, we take the 1D TFI model as an example. Near the QCP at $\tilde{\Delta}=0$, the isothermal Gr\"uneisen ratio $\Gamma_{\Delta}$ in Fig.~\ref{fg:1D_results}(b) exhibits a peak-dip structure. As the temperature decreases, the peak/dip values become sharper and the peak/dip locations approach the QCP. To analyze the divergent properties, we proceed to extract the scaling function from data collapse. By rescaling the axes as $x \equiv \tilde{\Delta} \tilde{T}^{-1/z\nu}$ and $y \equiv \Gamma_{\Delta} \tilde{T}^{1/z\nu}\Delta_c$, the Gr\"uneisen ratio data in Fig.~\ref{fg:1D_results}(b) exhibits excellent collapse on a single curve --- the scaling function $\phi(x)$, as shown in Fig.~\ref{fg:1D_results}(c). Similarly, Fig.~\ref{fg:1D_results}(d) illustrates that the Gr\"uneisen ratio can also be collapsed using $y \equiv \Gamma_{\Delta} \tilde{\Delta} \Delta_c$. The associated scaling function is $\tilde{\phi}(x)\equiv x\phi(x)$, as derived from Eq.~(\ref{fT},\ref{fg}). 

Particularly, for a self-dual QCP, the zero-order term of $\phi(x)$ vanishes [i.e., $\phi(0)=0$]. As shown in Fig.~\ref{fg:1D_results}(c,d), the $\phi(x)$ passes through the origin linearly $\phi(x\to 0) = \phi(0) + \phi'(0) x + O(x^2) \simeq \phi'(0) x$ and the $\tilde{\phi}(x)$ exhibits a quadratic behavior near $x=0$, i.e. $\tilde{\phi}(x\to 0) \simeq \phi'(0)x^2 $. Consequently, the Gr\"uneisen ratio remains non-divergent and temperature-independent above a self-dual QCP. While the Gr\"uneisen ratio itself does not fully capture the universal scaling behavior, it is important to note that the peak and dip values of the isothermal Gr\"uneisen ratio diverge following the universal scaling law, even for a self-dual QCP. According to Eq.~(\ref{fT}), the derivative of $\Gamma(T,g)$ with respect to $g$ is given by $\partial \Gamma(T,g)/\partial \tilde{g} = \tilde{T}^{-2/z\nu} / g_0 \cdot \phi'(x)$. The peak/dip conditions of the Gr\"uneisen ratio, determined by $\partial \Gamma(T,g)/\partial \tilde{g} = 0$, correspond to the extrema of scaling function $\phi(x)$, i.e.,  $\phi'(x)=0$. As illustrated in Fig.~\ref{fg:1D_results}(c), $\phi(x)$ exhibits both a peak and a dip, located at $x_{\rm peak}$ and $x_{\rm dip}$, respectively. For different temperatures, these extrema define two crossover lines described by $\tilde{g}=x_{\rm peak/dip} \cdot \tilde{T}^{1/z\nu}\propto {T}^{1/z\nu}$, which naturally delineate the boundaries of the QCR. Along the crossover lines, the peak/dip values of Gr\"uneisen ratio diverge following a universal scaling law $\Gamma_{\rm peak/dip}=\phi(x_{\rm peak/dip})/g_0 \cdot \tilde{T}^{-1/z\nu}\propto T^{-1/z\nu}$, according to the Eq.~\ref{fT}. Upon transitioning from the low-entropy regime into the QCR with strong thermal fluctuations, the significant entropy change gives rise to a quantum-critical-enhanced MCE with a universally diverging Gr\"uneisen ratio. 

In contrast to self-dual QCPs, the BEC QCP of 1D HAF model exhibits different universal behaivors. As shown in Fig.~\ref{fg:1D_results}(m,n), the isentropic lines reach their minima at locations displaced from the QCP, while the Gr\"uneisen ratio diverges as temperature decreases at $\tilde{h}=0$. The scaling function $\phi(x)$ still displays a peak-dip structure and $\tilde{\phi}(x)\equiv x \phi(x)$ passes through the origin [see Fig.~\ref{fg:1D_results}(o,p)]. Unlike the TFI model and $q$-state Potts model, $\phi(0)$ of the (1+2)D BEC QCP is a non-zero value and the Gr\"uneisen ratio is divergent above the QCP (i.e., $\tilde{h}=0$). 

Next, we move to the (2+1)D QCPs with linear dispersion. Fig.~\ref{fg:2D_results} shows the results of 2D TFI model, easy-plane $J_1$-$J_2$ dimer model and isotropic $J_1$-$J_2$ dimer model on a fixed system size $L=24$. As shown in Fig.~\ref{fg:2D_results}(a,e,i), at high temperatures, the Gr{\"u}neisen ratio exhibits a nearly flat curve and remains trivially positive. As temperature decreases, a peak progressively emerges and shifts toward the QCP. In the meantime, the 2D TFI model exhibits distinct sign change of Gr\"uneisen ratio near the QCP and the easy-plane $J_1$-$J_2$ dimer model also displays minor sign change, which is consistent with the theoretical prediction~\cite{PhysRevB.72.205129}. 

However, the Gr{\"u}neisen ratio in the 2D isotropic $J_1$-$J_2$ dimer model is always positive near the QCP, as shown in Fig.~\ref{fg:2D_results}(i). This anomalous behavior originates from the low-temperature thermal entropy of gapless excitations. For the gapped dimer phase, a finite energy gap $\Delta_E\sim K-K_c$ emerges and the entropy decays exponentially as $s(K>K_c)\propto e^{-\Delta_E/T}$. At the (2+1)D O(3) QCP, quantum critical fluctuations take a power-law decaying entropy $s(K_c)\propto T^{d/z} \propto T^2$, determined by the spatial dimension $d$ and the dynamic exponent $z$ ($d=2$ and $z=1$ for this case). In contrast, within the Néel phase, the breaking of SU(2) symmetry gives rise to two Goldstone modes, which also produce a quadratic entropy contribution, $s(K<K_c) \propto T^2$, comparable to that at the QCP. Therefore, the Gr\"uneisen ratio, which is related to the isothermal entropy change $(\partial s/\partial K)_T$, may not change sign with the external field $K$, since the low-temperature entropy above the QCP is not maximal at fixed temperatures. On the other hand, the Gr{\"u}neisen ratio, which characterizes the adiabatic cooling processes, encodes information about isentropes. According to Eq.~(\ref{gamma}), the isentropic lines can be obtained differentially as $T_f = T_i + \delta T = T_i + T \cdot \Gamma_g \cdot \delta g$ [see Fig.~\ref{fg:2D_results}(b,f,j)]. As we can see, unlike conventional cases, the temperature decreases continuously during the demagnetization process rather than exhibits a valley near the QCP. 
This feature could be beneficial for cooling because it means that after cooling by reducing the external field from a large value to $g_c$, the coolant can hold its temperature even when the field is further reduced.

On a finite lattice, the inverse temperature $\beta$ acts as a cutoff. When $\beta$ is much smaller than $L$, the scaling functions can be effectively approximated as univariate functions $\phi(x)$ and $\tilde{\phi}(x)$ in the thermodynamic limit (see Appendix~\ref{append:convergence}), which are sufficient to guide and benchmark experiments. We obtain scaling functions $\phi(x)$ and $\tilde{\phi}(x)$ for 2D quantum systems through data collapse, as shown in Fig.~\ref{fg:2D_results}(c,d,g,h,k,l). Using the known critical exponents directly, the obtained data nearly fall on a single curve with only minor deviations at small $\beta$. These scaling functions are distinct across different universality classes. The behaviors of  scaling functions for the Gr{\"u}neisen ratio in the 2D spin models essentially coincide with those of the 1D spin models without self-duality. However, as illustrated in Fig.~\ref{fg:2D_results}(k,l), the scaling function $\phi(x)$ and $\tilde{\phi}(x)$ for 2D isotropic $J_1$-$J_2$ dimer model near the O(3) QCP is  different from the other two models. Here, $\phi(x)$ always remains positive and $\tilde{\phi}(x)$ does not exhibit a local minimum.

\section{Discussions}
\label{section5}
In this paper, we investigate several representative spin models with generic symmetry-breaking QCPs to extract universal scaling functions of Gr{\"u}neisen ratio from numerical data. For 1D spin models, our results provide the scaling functions in the thermodynamic limit. For 2D spin models, we study the scaling function of the Gr{\"u}neisen ratio on finite-size lattice with a maximum system size of $L=24$. We construct a systematic library of scaling functions that encompasses diverse universality classes. Our work provides numerical diagnosises for the Gr{\"u}neisen ratio and we hope it will motivate and guide more MCE measurements near the QCPs in novel quantum materials.  

Additionally, we analyzed two special cases: the self-dual QCP and (2+1)D O(3) QCP. The Gr{\"u}neisen ratio in 1D TFI and quantum $q$-state Potts models remains finite at QCP, owing to the constraint imposed by self-duality. The corresponding scaling function exhibits distinct behavior from 1D HAF model, where $\phi(x)$ always passes through the origin and $\tilde{\phi}(x)\sim x^2$ near $x=0$. For the isotropic $J_1$-$J_2$ dimer model, we observed that the sign of the Gr{\"u}neisen ratio near the O(3) QCP remains positive and the obtained scaling functions exhibit distinct characteristics, which may indicate better cooling performance. This furthuer motivates the study of scaling behavior of the Gr{\"u}neisen ratio near O(N) Wilson-Fisher QCP (N>3). 

Since the results in 2D models are affected by finite-size effects, it would be desirable to derive more precise results through conformal perturbation theory~\cite{James_2018,PhysRevD.91.025005}. Moreover, it is worthwhile to build alternative models with small finite-size effects via fuzzy sphere regularization~\cite{PhysRevX.13.021009}, which has been implemented for studying 3D Ising conformal field theory (CFT)~\cite{PhysRevX.13.021009,PhysRevLett.131.031601}, 3D XY CFT~\cite{Lauchli2017} and 3D O(3) CFT~\cite{PhysRevB.110.115113}. 

On the other hand, the Gr{\"u}neisen ratio also serves as a powerful probe for QCPs, it would be valuable to extend our computation to other sophisticated quantum materials with unconventional QCPs, for instance, diagnosing deconfined quantum critical point (DQCP)~\cite{doi:10.1126/science.1091806,PhysRevB.70.144407,PhysRevB.100.125137,Xi_2022} in Shastry-Sutherland lattice compound \ch{SrCu_2(BO_3)_2}~\cite{doi:10.1126/science.adc9487} or Gross-Neveu QCP~\cite{PhysRevD.10.3235,PhysRevB.106.075111} in twisted bilayer graphene~\cite{Huang_2025}. 

\begin{acknowledgements}
We acknowledge the support from National Key R\&D Program of China (Grant No.2022YFA1403400) and from NSFC (Grant No. 12374144). This work was supported by the National Key Projects for Research and Development of China (Grant No.~2024YFA1409200), the National Natural Science Foundation of China (Grant Nos.~12222412, 12447101), and the Strategic Priority Research Program of Chinese Academy of Sciences (Grant No. XDB1270000). We thank the Beijing PARATERA Tech CO.,Ltd. (URL:https://cloud.paratera.com) and the HPC-ITP for providing HPC resources and generous allocation of CPU time that have contributed to the research results reported within this paper. 

\end{acknowledgements}

\appendix

\section{UNIVERSAL DIVERGENT SCALING OF THE GR{\"U}NEISEN RATIO }
\label{append:scaling}
The divergent scaling behavior of Gr{\"u}neisen ratio is a universal thermodynamic signature of QCP, which can be derived from the hyperscaling ansatz of the singular part of free energy density~\cite{Zhang_2023} 
\begin{equation}
\begin{aligned}
f_s(T,g,L) &= -f_0\tilde{T}^{(d+z)/z} \cdot F\left( {\tilde{g}}{\tilde{T}^{-1/z\nu}},{\tilde{T}^{-1/z}L^{-1}}\right) \\
&=-f_0 \tilde{g}^{\nu(d+z)} \cdot \widetilde{F}\left({\tilde{g}}{\tilde{T}^{-1/z\nu}},{\tilde{T}^{-1/z}L^{-1}}\right),
\label{fs}
\end{aligned}
\end{equation}
where $f_0$ is a nonuniversal constant. $F(x,y)$ and $\widetilde{F}(x,y) \equiv x^{-\nu(d+z)}F(x,y)$ are hyperscaling functions with $x\equiv \tilde{g}\tilde{T}^{-1/z\nu}$ and $y\equiv \tilde{T}^{-1/z} L^{-1}$. The variable $y$ approaches 0 in the thermodynamic limit $L\to \infty$ and thus the bivariate function $F(x,y)$ and $\widetilde{F}(x,y)$ reduce to the univariate form in Refs.~\cite{PhysRevLett.123.230601,PhysRevLett.91.066404}. Following the derivation in Ref.~\cite{PhysRevLett.91.066404}, we can derive the finite-size scaling forms of the Gr{\"u}neisen ratio,
\begin{equation}
\Gamma(T,g,L)= \tilde{T}^{-1/z\nu}/g_0\cdot \Phi\left( x,y\right),
\label{fT2}
\end{equation}
\begin{equation}
\Gamma(T,g,L)= \tilde{g}^{-1}/g_0\cdot \widetilde{\Phi}\left( x,y\right),
\label{fg2}
\end{equation}
where $\Phi(x,y)$ and $\widetilde{\Phi}(x,y) \equiv x\Phi(x,y)$ are universal finite-size scaling functions of Gr{\"u}neisen ratio. Different from the general thermodynamical quantities, the Gr\"uneisen ratio is independent of nonuniversal constant $f_0$. For the infinite-size spin model ($L\to \infty$, thus $\tilde{T}^{-1/z}L^{-1} \to 0$), the bivariate scaling functions $\Phi(x,y)$ and $\widetilde{\Phi}(x,y)$ become univariate functions $\phi(x)\equiv\Phi(x,0)$ and $\tilde{\phi}(x)\equiv\tilde{\Phi}(x,0)$. 

 Generally, $\phi(x)$ can be expanded as $\phi(x) = \phi(0) + \phi'(0)x + \phi''(0)x^2/2 + \mathcal{O}(x^3)$ near $x=0$, and the corresponding $\tilde{\phi}(x)$ always passes through the origin and can be expanded as $\tilde{\phi}(x) = \phi(0)x + \phi'(0)x^2 + \phi''(0)x^3/2 + \mathcal{O}(x^4)$. Furthermore, within the language of scaling functions, we can derive the quantum critical scalings, 
\begin{equation}
\Gamma_g \propto {T}^{-1/z\nu}, ~~~~~~~~~~~~\tilde{T}\gg \vert \tilde{g} \vert^{z\nu},
\label{eq:Tscaling}
\end{equation}
\begin{equation}
\Gamma_g \propto {(g-g_c)}^{-1}, ~~~~~~\tilde{T}\ll \vert \tilde{g} \vert^{z\nu},
\label{eq:Bscaling}
\end{equation}
where $\tilde{T} \sim \vert \tilde{g} \vert^{z\nu}$ denotes crossovers. For $\tilde{T} \gg |\tilde{g}|^{z\nu}$ (i.e., $x\ll 1$), $\Gamma_g \approx \tilde{T}^{-1/z\nu}/g_0 \cdot \phi(0) \propto {T}^{-1/z\nu}$. On the other hand, for $\tilde{T} \ll |\tilde{g}|^{z\nu}$ (i.e., $x\gg 1$), $\Gamma_g \approx \tilde{g}^{-1}/g_0 \cdot \tilde{\phi}(\infty) \propto (g-g_c)^{-1}$, where $\tilde{\phi}(\infty)$ is the function value of $x\to \infty$. In particular, for a self-dual QCP, the prefactor vanishes [i.e., $\phi(0)=0$] and the Gr{\"u}neisen ratio remains finite for $\tilde{T}\gg \vert \tilde{g} \vert^{z\nu}$~\cite{PhysRevLett.123.230601,PhysRevB.97.245127,Zhang_2023}.

\section{FINITE SIZE SCALING FUNCTIONS AND CRITICAL EXPONENTS IN DIFFERENT UNIVERSALITY CLASSES}
\label{append:finite}
\begin{figure}[htp!]
    \centering
    \includegraphics[width=1.0\columnwidth]{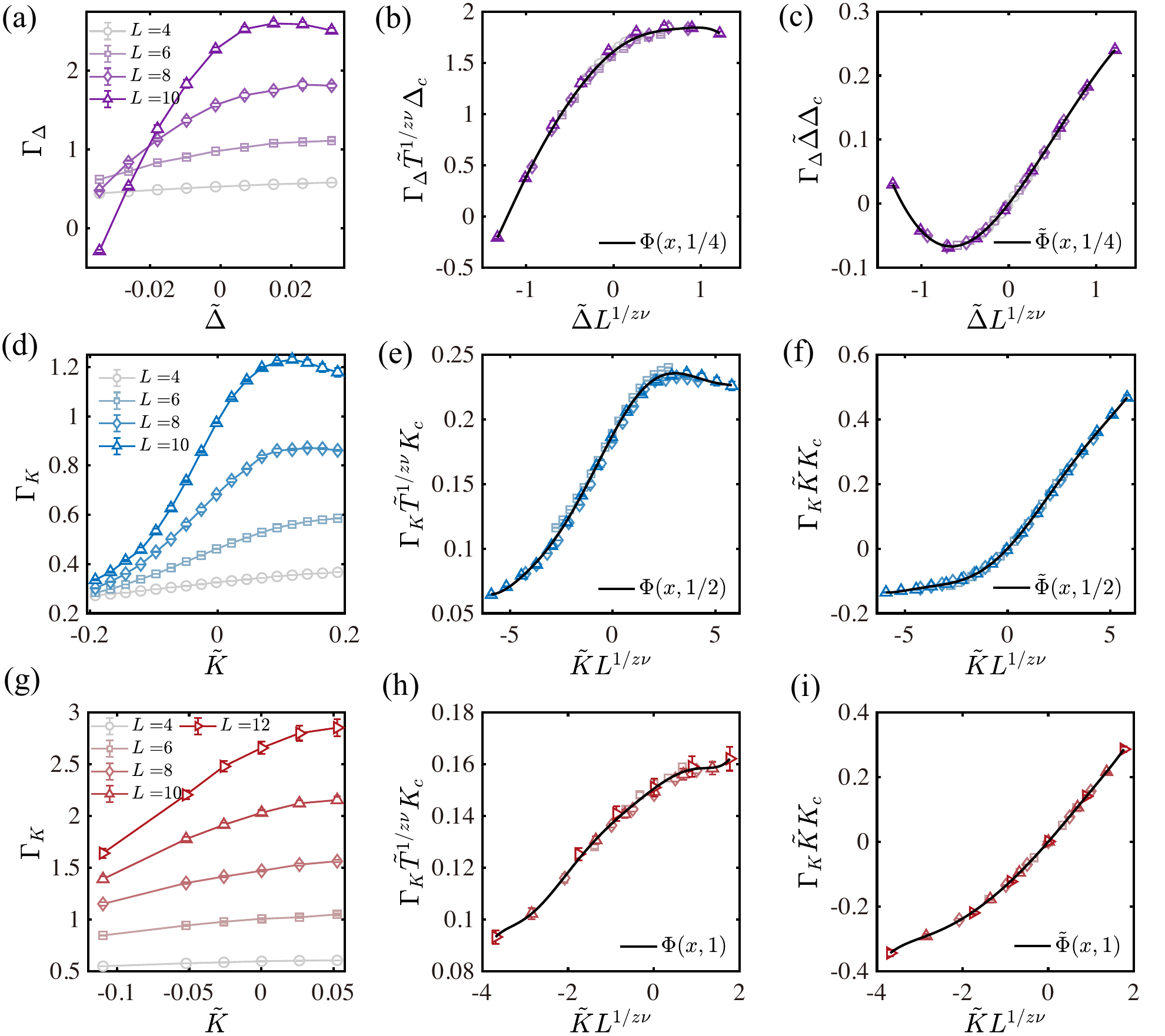}
    \caption{\textbf{Gr{\"u}neisen ratio $\Gamma_g$ calculated for 2D spin models on finite-size lattices with fixed $y$.} (a), (d) and (g) Gr{\"u}neisen ratio as function of tune parameter for different system size $L$ for the 2D TFI model, 2D easy plane $J_1$-$J_2$  model and 2D isotropic $J_1$-$J_2$ model. (b), (e) and (h) The data collapse of $\Gamma_g$ near the QCP according to Eq.~(\ref{fT}). Here we choose $T_0=1$, $g_0=g_c$ and use the (2+1)D Ising critical exponent $\nu=0.629971$, (2+1)D XY critical exponent $\nu=0.67175$ and (2+1)D O(3) critical exponent $\nu=0.707$ in the scaling sequentially. (c), (f) and (i) The data collapse of $\Gamma_g$ near the QCP according to Eq.~(\ref{fg}).}
    \label{fg:2D_beta-L}
\end{figure}

In the main text, we investigate the universal scaling functions of the Gr{\"u}neisen ratio approaching the thermodynamic limit. Under finite-size conditions, the Gr{\"u}neisen ratio may exhibit distinct scaling behaviors, e.g., the Gr{\"u}neisen ratio in 1D TFI model with periodic boundary condition diverges at QCP for fixed nonzero $y$~\cite{Zhang_2023}.  Here we turn to the finite-size scaling function where $\beta$ scales with system size $L$. In this scenario, we still can get a univariate scaling function. As is illustrated in Fig.~\ref{fg:2D_beta-L}, we calculate the Gr{\"u}neisen ratio for above 2D spin models in the case of $y=\text{const}$. In Fig.~\ref{fg:2D_beta-L}(a,d,g), the Gr{\"u}neisen ratio diverges at QCP as expected. Here we take $T_0=1$ and the variable $y$ in scaling function is 1 for 2D isotropic $J_1$-$J_2$ model. While we take $y=1/4$ in the 2D TFI model and $y=1/2$ in the 2D easy plane $J_1$-$J_2$ model due to the specific heat in this model is relatively small and fluctuates strongly at low temperatures and in high external fields. The data collapse method follows the same approach as that for the 1D model, but the horizontal axis has been rescaled to $\tilde{g}L^{1/z\nu}$ in accordance with the convention of finite-size scaling laws. The corresponding critical exponent is used in the data collapse according to the QPT universality class. Theoretical predictions are confirmed with SSE QMC simulations, which means the universal divergence of the Gr{\"u}neisen ratio also manifest itself on the finite-size lattices.

The Gr{\"u}neisen ratio also provides a means to extract the critical exponents numerically. Using the known critical exponents from universality classes already yields reasonably good data collapse. In Table~\ref{tab:exponents}, we show the results of fitting $\nu$ for (2+1)D QCPs according to Eq.~(\ref{fg}). We use a polynomial of 5th order for fitting scaling function from QMC data at fixed $y$. The fitted value of $\nu$ is already consistent with the reference value on small lattices.
\begin{table}[htp!]
	\centering
	\begin{tabular}{ccccc}
		\toprule
		Universality class & $d$ & $z$ & $\nu$ & Fit $\nu$ \\ 
        \midrule
		Ising & 2 & 1 & 0.629971 & 0.64(2)\\
        \midrule
		XY & 2 & 1 & 0.67175 & 0.68(2)\\
        \midrule
		O(3)& 2 & 1 & 0.707 & 0.71(2)\\
		\bottomrule
	\end{tabular}
    \caption{Critical exponents for different universality classes.}
    \label{tab:exponents}
\end{table}

\section{CONVERGENCE OF QMC DATA ON FINITE SIZE LATTICE}
\label{append:convergence}
Here we compute the extrapolation of $\Gamma_g$ with respect to $L$ at fixed $\beta$, thereby establishing a criterion for data convergence. 

\begin{figure}[htp!]
    \centering
    \includegraphics[width=1.0\columnwidth]{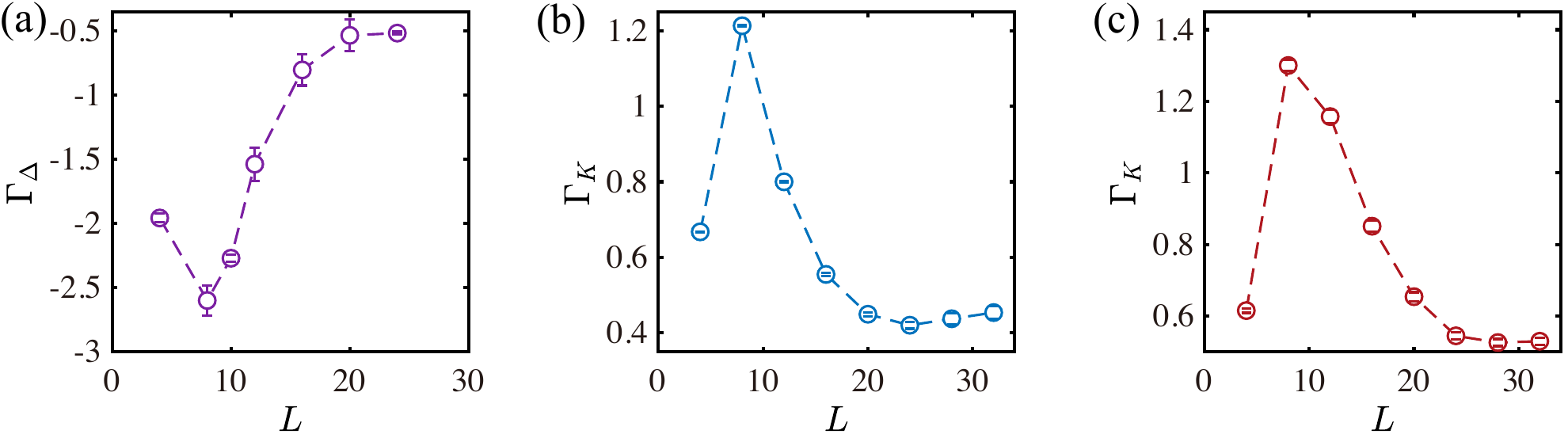}
    \caption{\textbf{Gr{\"u}neisen ratio calculated for 2D spin models near QCP as function of $L$.} (a) 2D TFI model at $\Delta=3.04$ and $\beta=2.5$. (b) Easy plane $J_1$-$J_2$ model  at $K=2.10$ and $\beta=5.0$. (c) isotropic $J_1$-$J_2$ model at $K=1.91$ and $\beta=6.0$.}
    \label{fg:gr-L}
\end{figure}

Taking the isotropic $J_1$-$J_2$ model as an example, Fig.~\ref{fg:gr-L}(c) shows that when $\beta=6.0$, $\Gamma_K$ gradually converges with increasing $L$ and stabilizes when $L>20$. For smaller $\beta$, a lattice size of $L=24$ is generally sufficient to obtain a convergent result. 

For fixed system size $L$, $y \equiv 1/(\tilde{T}L)$ in the scaling function can not be a constant, data collapse is influenced by strong finite-size effects. In experiments, the thermodynamic limit $L\to\infty$ leads to $1/(\tilde{T}L)\to0$, thus guaranteeing a well-defined univariate scaling function. This is challenging to achieve in QMC simulations. However, we argue that when $\beta$ is not excessively large but sufficiently close to the QCP, allowing $\Gamma_g$ to satisfy scaling law, a fixed $L$ is adequate to obtain converged data. For $\beta$ too small (far from QCP), the scaling form fails. On the other hand, if $\beta$ is too large, the current system size $L$ may not guarantee sufficiently converged data. Thus, data collapse is only achievable in a limited temperature window. Certainly, a larger system size would give more precise results and  better data collapse.

To obtain reliable results at lower temperatures, a simultaneous increase in system size $L$ is required, yet this imposes greater computational demands on the SSE QMC method, whose computational complexity scales as $\mathcal{O}(\beta N)$.

\section{EXPRESSIONS OF SCALING FUNCTIONS}
\label{append:functions}
We make a table of polynomial fits for universal scaling function $\phi(x)$  for experimental researchers to consult and compare. Analytic solutions for the 1D TFI model and HAF model are provided in Refs.~\cite{PhysRevB.97.245127,doi:10.1126/sciadv.aao3773}.

\begin{table*}[htp!]
    \centering
    \renewcommand{\arraystretch}{1.4}
    \begin{tabularx}{\textwidth}{cXc}
    \toprule
     Universality class  & Universal scaling function $\phi(x)$ & Range of $x$   \\
     \hline
     (1+1)D Ising & $-1.23\times10^{-5} x^{15} - 9.78\times10^{-8}x^{14} + 4.52\times10^{-4}x^{13} + 3.39\times 10^{-6}x^{12} - 6.84\times 10^{-3} x^{11} - 4.77\times 10^{-5} x^{10} + 5.52\times10^{-2}x^{9} + 3.51\times 10^{-4}x^{8} - 0.26 x^{7} - 1.46\times10^{-3}x^{6} + 0.70x^{5} + 3.46\times 10^{-3}x^{4} - 1.13x^{3} - 4.78\times10^{-3}x^{2} + 1.14x$ & [-2,2] \\
     (1+1)D 3-state Potts & $-8.92\times10^{-3}x^{15} + 3.19\times10^{-4}x^{14} + 1.45\times 10^{-1}x^{13} - 4.76\times 10^{-3}x^{12} - 9.63\times 10^{-1}x^{11} + 2.85\times 10^{-2}x^{10} + 3.41 x^9 - 8.77\times 10^{-2}x^8 - 6.93x^7 + 1.47\times 10^{-1}x^6 + 8.22x^5 - 1.29\times 10^{-1}x^4 - 5.67x^3 + 5.02\times 10^{-2}x^2 + 2.36x $ & [-2,2] \\
     (1+1)D 4-state Potts & $-1.94\times 10^{-2}x^{15} + 3.40\times 10^{-3}x^{14} + 3.11\times 10^{-1}x^{13} - 5.03\times 10^{-2}x^{12} - 2.04x^{11} + 2.99\times 10^{-1}x^{10} + 7.05x^9 - 9.10\times 10^{-1}x^8 - 1.38\times 10^1 x^7 + 1.50x^6 + 1.55\times 10^1 x^5 - 1.30x^4 - 9.56x^3 + 5.14\times 10^{-1}x^2 + 3.13x $ & [-2,2] \\
     (1+2)D BEC & $ 7.78\times 10^{-4}x^{15} - 2.25\times 10^{-3}x^{14} - 1.23\times 10^{-2}x^{13} + 3.52\times 10^{-2}x^{12} + 7.81\times 10^{-2}x^{11} - 2.25\times 10^{-1}x^{10} - 2.55\times 10^{-1}x^9 + 7.62\times 10^{-1}x^8 + 4.40\times 10^{-1}x^7 - 1.49x^6 - 3.26\times 10^{-1}x^5 + 1.71x^4 - 1.05\times 10^{-1}x^3 - 1.18x^2 + 4.18\times 10^{-1}x + 5.05\times 10^{-1} $ & [-2,2] \\
     (2+1)D Ising & $-1.2587x^5+1.4687x^4+0.2083x^3-1.0636x^2+0.5919x+0.1175$ & [-0.65, 0.79] \\ 
     (2+1)D XY & $-2\times10^{-6}x^7-7.62\times10^{-6}x^6+9.53\times10^{-5}x^5+2.37\times10^{-4}x^4-0.002x^3-0.002x^2+0.0241x+0.0332$ & [-5.45, 4] \\
     (2+1)D O(3) & $-1.56\times10^{-6}x^7+3.425\times10^{-6}x^6+8.46\times10^{-5}x^5-1.41\times10^{-4}x^4-0.00184x^3+0.0019x^2+0.0229x+0.035$ & [-4.37, 5.11] \\
      
    \bottomrule
    \end{tabularx}
    \caption{Polynomial fitting expressions of universal scaling functions.}
    \label{tab:functions}
\end{table*}

\section{NUMERICAL METHODS}
In this section, we give the complete details of the numerical methods used in this work for calculating the Gr{\"u}neisen ratio. 
\label{append:method}
\subsection{Thermal Tensor Network}
The linearized tensor renormalization group (LTRG) method is employed to simulate the infinite-size 1D systems~\cite{Li2011,Dong2017,Chen2018,Lih2019,tanTRG2023}. To ensure high numerical convergence, we push the bond dimension up to $D=500$. For a thermal density matrix at inverse temperature $\beta\equiv 1/T$, we take Trotter-Suzuki decomposition as follows, 
\begin{equation}
    \rho(\beta) = e^{-\beta H} = e^{-\beta (H_1+H_2)} \approx \left(e^{-\tau H_1}e^{-\tau H_2}\right)^n,
\end{equation}
where $H_1$ ($H_2$) is the local Hamiltonian living on even (odd) bonds, and $\tau=\beta/n$ is the Trotter step. Then, the partition function can be obtained by the bilayer trace~\cite{Dong2017}, 
$Z(\beta) =  {\mathrm{Tr}} [\rho(\beta/2) \rho(\beta/2)^\dagger]$, and the free energy density reads 
\begin{equation}
    f = -\frac{1}{\beta} \ln Z(\beta).
\end{equation}
Consequently, thermodynamics quantities such as generalized thermal expansion $\alpha_g$ 
and specific heat $c_g$ can be obtained as, 
\begin{equation}
    \alpha_g = \frac{\partial^2 f}{\partial T \partial g} ,\quad c_g = - T \frac{\partial^2 f}{\partial T^2}.
\end{equation}

\subsection{Quantum Monte Carlo}
For 2D spin models, we employ the stochastic series expansion quantum Monte Carlo (SSE QMC) method \cite{Sandvik_2010,sandvik2019stochastic,AWSandvik_1992} based on the loop algorithm \cite{PhysRevE.68.056701,PhysRevB.59.R14157,PhysRevE.66.046701}. In this  subsection, we introduce the measurement of the Gr{\"u}neisen ratio $\Gamma_g$ in SSE QMC simulations. The partition function can be Taylor expanded as
\begin{equation}
Z=\text{Tr} \left\{ e^{-\beta H} \right\}=\sum_\Psi\sum_{n=0}^{\infty}\frac{\beta^n}{n!}\langle\Psi\vert(-H)^n\vert\Psi\rangle,
\end{equation}
where $\{\vert \Psi\rangle\}=\{\vert S_1^z,S_2^z,\cdots,S_N^z\}$ is a complete basis of an N-site quantum spin system. Considering a Hamiltonian $H(g)=H_1+g H_2$ with $g$ the external field, we decompose it into bond operators 
\begin{equation}
H=-\sum_{a,b}H_{a,b},
\end{equation}
where $a\in\{1,2\}$ denotes a diagonal or off-diagonal operator in Hamiltonian and $b$ denotes bond index for a pair of sites $i(b),j(b)$. In practice, $n$
is truncated at a fixed length $M$ by introducing $M-n$ unit operators $H_{0,0}=I$ in all possible ways. The partition function can then be expressed as
\begin{equation}
Z=\sum_\Psi \sum_{S_M}\frac{\beta^n(M-n)!}{M!}\langle\Psi\vert\prod_{i=1}^M H_{a_i,b_i}\vert\Psi\rangle,
\end{equation}
where $S_M$ denotes the operator string and $n$ is the number of non-unit bond operators in it. The specific heat is calculated from the variance of the energy. In SSE QMC, one can show that specific heat of a finite-size lattice is
\begin{equation}
c_g=\frac{\langle n^2\rangle-\langle n\rangle-\langle n\rangle^2}{N}.
\end{equation}

The generalized thermal expansion is given as $\alpha_g=(1/N)\left(\partial H_2/\partial T \right)_g$. And then it can be written as
\begin{equation}
\alpha_g=\frac{1}{N}\frac{\langle H_2 H \rangle-\langle H_2\rangle\langle  H \rangle}{T^2}.
\label{alpha}
\end{equation}
For any operator $H_{a,b}$ in the Hamiltonian, diagonal or off-diagonal, its expectation value is simply given in Ref. \cite{sandvik2019stochastic,AWSandvik_1992} as $\langle H_{a,b}\rangle = -\langle n_{a,b}\rangle/\beta$. Summing over all $a,b$ belonging to the corresponding observables, each term in Eq.~(\ref{alpha}) is given as
\begin{equation}
\langle  H\rangle=-\frac{\langle n\rangle}{\beta},
\end{equation}

\begin{equation}
\langle H_2\rangle=-\frac{\langle n_2\rangle}{\beta g},
\end{equation}

\begin{equation}
\langle H_2 H\rangle=\frac{\langle (n-1)n_2\rangle}{\beta^2 g},
\end{equation}
where $n_2$ is the number of bond operators in $H_2$ in the sampled operator string $S_M$. Thus we can measure $\Gamma_g$ in SSE QMC simulations. We use 96 parallel processes to obtain statistically independent measurements of observables. Each process performs $10^5$ Monte Carlo steps for thermalization and $10^5$ measurement steps per bin.

\bibliography{ref}

\end{document}